# Layer-dependent magnetic property in a superconducting quintuple-layer nickelate La$_6$Ni$_5$O$_{12}$

Terri Yoon[1,2] and Myung Joon Han[1,*]

[1] *Department of Physics, KAIST, Daejeon 34141, Republic of Korea*
[2] *Department of Aerospace Engineering, Republic of Korea Air Force Academy*

To investigate the detailed magnetic properties of a recently discovered superconducting nickelate Nd$_6$Ni$_5$O$_{12}$, we performed the first-principles electronic structure calculation based on density functional theory. The band dispersion, electronic charge distribution and the magnetic moment are computed with La substituted for Nd, and compared with another structural type of nickel-based superconducting material, namely, RNiO$_2$ (R: rare-earth elements). In particular, we estimated the magnetic exchange interaction strength based on magnetic force theory. Our results show that the inter-atomic magnetic couplings are notably reduced by intrinsic hole doping from the blocking fluorite slab which validates the conventional view of regarding Nd$_6$Ni$_5$O$_{12}$ as a doped case of its infinite-layer counterpart. At the same time, however, the interactions are markedly layer-dependent. The outer most NiO$_2$ layer adjacent to the blocking fluorite roughly corresponds to the 20% chemical doping in the infinite-layer material whereas the inner layers have stronger couplings. The long-range nature and the out-of-plane interactions are also presented. Our results provide useful information to understand this new superconducting nickelate whose intrinsic layer structure is obviously distinctive.

## I. INTRODUCTION

Ever since the first discovery of copper oxide high critical temperature (T$c$) superconductor in 1986, tremendous theoretical and experimental efforts have been devoted to understand its nature and the intriguing relationships with the nearby phases [1-10]. Even after the more than thirty years intensive investigations, this material and the related issues are still in the central place of condensed matter physics. An interesting research direction has been to find or design the unconventional superconductivity in an alternative material platform, namely, non-copper-based compounds, which can help us to understand the exotic superconductivity by scrutinizing their similarities and differences [11-27]. Several design principles have been highlighted in this direction based on the commonalities.

In this respect, nickelates have attracted significant attention [11-23]. The long-sought goal has recently been achieved by Li et al. who successfully synthesize a thin film of infinite-layer nickelate Nd$_{1-x}$Sr$_x$NiO$_2$ [28]. Originally at the doping level of x=0.2 and then around this range, superconductivity was observed with T$c$ ~10-15 K [28, 29]. While magnetic interactions and fluctuations are deemed crucially important for pairing, several other characteristics have been featured in regard to its unconventional natures of normal as well as superconducting phase [11-23,28-45].

Notably, this first discovery is soon after followed by another. Pan et. al. reported the second case on nickelate superconductor with a distinctive structural type [46]. Contrary to its infinite layer cousin, this new material Nd$_6$Ni$_5$O$_{12}$ (Nd6512) becomes superconducting without chemical doping. Owing to its intrinsically distinctive feature, namely the quintuple layer structure, additional question arise to understand the superconductivity and the possibly existing nearby phases such as pair density wave state [9]. Previous electronic structure calculation studies reported the useful standard information focusing on the difference between infinite- and quintuple-layer material at the level of density functional theory (DFT) [54, 55], DFT+U [54] and DFT+DMFT (dynamical mean-field theory) [47,55]. However, given that magnetism is of key importance also in this new family, the detailed information about the magnetic interactions in this intrinsically and obviously different structure needs to be carefully examined and compared with the related materials. Whereas several theoretical studies investigated the magnetic interaction profile for the infinite layer $RNiO_2$ (R112; R: rare-earth ions such as Nd, La, and Pr) [50,52,53], the corresponding information is still lacking for the new quintuple layer nickelate.

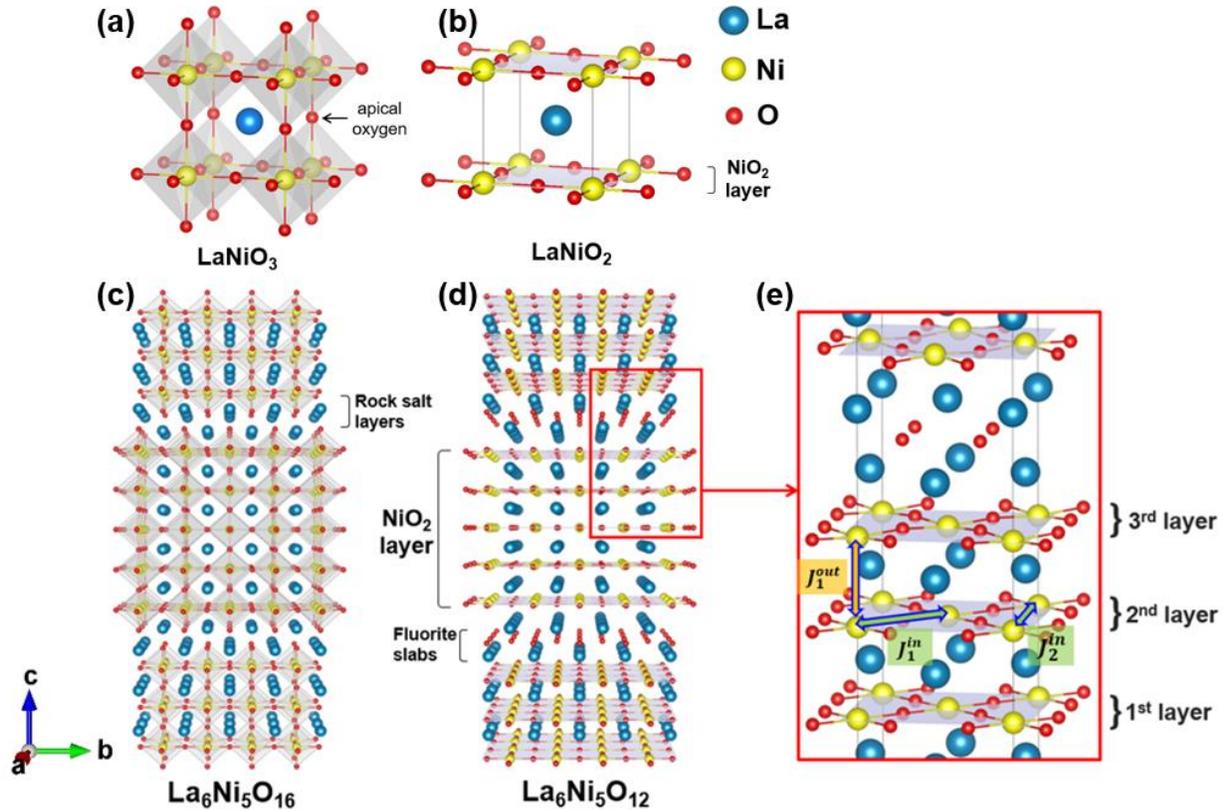

**Figure 1.** Crystal structures of the infinite- and the quintuple-layer nickelates. (a) $RNiO_3$, (b) $RNiO_2$, (c) $R_6Ni_5O_{16}$ and (d) $R_6Ni_5O_{12}$. In quintuple-layer nickelate R6512, five layers of $NiO_2$ are separated by fluorite blocking slabs as shown in (d). Turquoise, yellow, and red spheres denote rare-earth, Ni, and O atoms, respectively. The enlarged picture of R6512 is presented in (e). For convenience, the middle layer (among the five) of quintuple-layer nickelate is designated as the 1st layer. The adjacent layer to fluorite is the third, and the second $NiO_2$ layer is in between the 1st and the 3rd as depicted in (e). The inter-atomic magnetic interaction strengths within the plane and along the out-plane direction are denoted by $J^{in}$ and $J^{out}$, respectively. The subscript $n$ refers to the $n$-th neighbor coupling.

In this paper we performed the detailed investigation of magnetic interactions within and out-of-plane $NiO_2$ planes based on the first-principles density functional theory combined with force response theory. As indicated by the nominal valence and reported in previous investigations [46,55], the charge transfer from blocking fluorite slabs onto $NiO_2$ makes it corresponding to the hole-doped infinite layer case with the doping level of $\delta=0.2$. At the same time, however, the charge redistribution is layer-dependent. We found that the magnetic interactions are also clearly layer-dependent. As one moves from the inner to the outer layer of quintuple structure, the major in-plane coupling gets gradually reduced, and this feature is maintained over the long range. Our results show how the charge imbalance across the layers affects the magnetic profile, and thereby provide an important piece of information to understand new quintuple-layer superconductivity in comparison with other related materials.

## II. COMPUTATIONAL METHODS

DFT calculations were carried out within the Perdew-Burke-Ernzerhof parametrization of generalized gradient approximation (GGA) as implemented in VASP (Vienna ab initio simulation package) [48]. We considered $La_6Ni_5O_{12}$ (La6512) instead of $Nd_6Ni_5O_{12}$ (Nd6512) to avoid the numerical instabilities arising from Nd-$f$ electron following the previous studies [50,52-56]. The optimized lattice parameters of Ref.[54], k-grid mesh of 10×10×10, and the energy cutoff of 500 eV were used in our calculation. To simulate charge doping, we employed virtual crystal approximation (VCA) for the pseudopotential of La, which is likely more important in this case of quintuple-layer system because the fluorite blocking layer ($R_2O_2$) is known to be a hole provider for the $NiO_2$ planes. For comparison, we also calculated the 112 compound for which the same computation setup was adopted as in Ref.50. $La_6Ni_5O_{16}$ is also computed for which the relaxed lattice parameters were taken from the previous study [57], and 8×8×2 k-mesh and 500 eV energy cutoff were used.

To obtain exchange coupling constants, we performed magnetic force theory (MFT) calculations [58,59] by using our $J_x$ code [60-62]. Here, the magnetic exchange coupling constant $J_{ij}$ between atom $i$ and $j$ is computed based on magnetic force linear-response theory [58,59,61,62]. The following convention for the spin model Hamiltonian was used:

$$H = -\sum_{i \neq j} J_{ij} \boldsymbol{e_i} \cdot \boldsymbol{e_j} \qquad (1)$$

where $\boldsymbol{e}_{i,j}$ refers to the spin unit vector of atomic sites $i$ and $j$. Antiferromagnetic (AFM) and ferromagnetic (FM) coupling is expressed as positive and negative sign, respectively.

## III. RESULTS AND DISCUSSION

Fig. 1(a) and (c) shows the crystal structure of two members of $R_{n+1}Ni_nO_{3n+1}$ Rudlessden Popper (RP) series (n=2, 3, ..., ∞), namely, $RNiO_3$ (n=∞) and $R_6Ni_5O_{16}$ (n=5), respectively. Apical oxygen reduction leads them to be $RNiO_2$ (R112; see Fig.1(b)) and $R_6Ni_5O_{12}$ (R6512; see Fig.1(d)), respectively, belonging to series of $R_{n+1}Ni_nO_{2n+2}$ [54]. The removal of apical oxygens converts the octahedral coordination around Ni ions to be square-planar. While R112 is composed of quasi-two-dimensional $NiO_2$ layers connected infinitely along the out-of-plane direction, in tetragonal R6512 (space group $I4/mmm$), quintuple layers of $NiO_2$ are separated from each other by the blocking 'fluorite slabs' [46], and the five-layer bundle is shifted by a half of the in-plane lattice constants. Fig. 1(e) is an enlarged picture in which we designate the layer numbers and the magnetic coupling constants. It is in these two compounds of R112 and R6512 that superconductivity was reported [28,46].

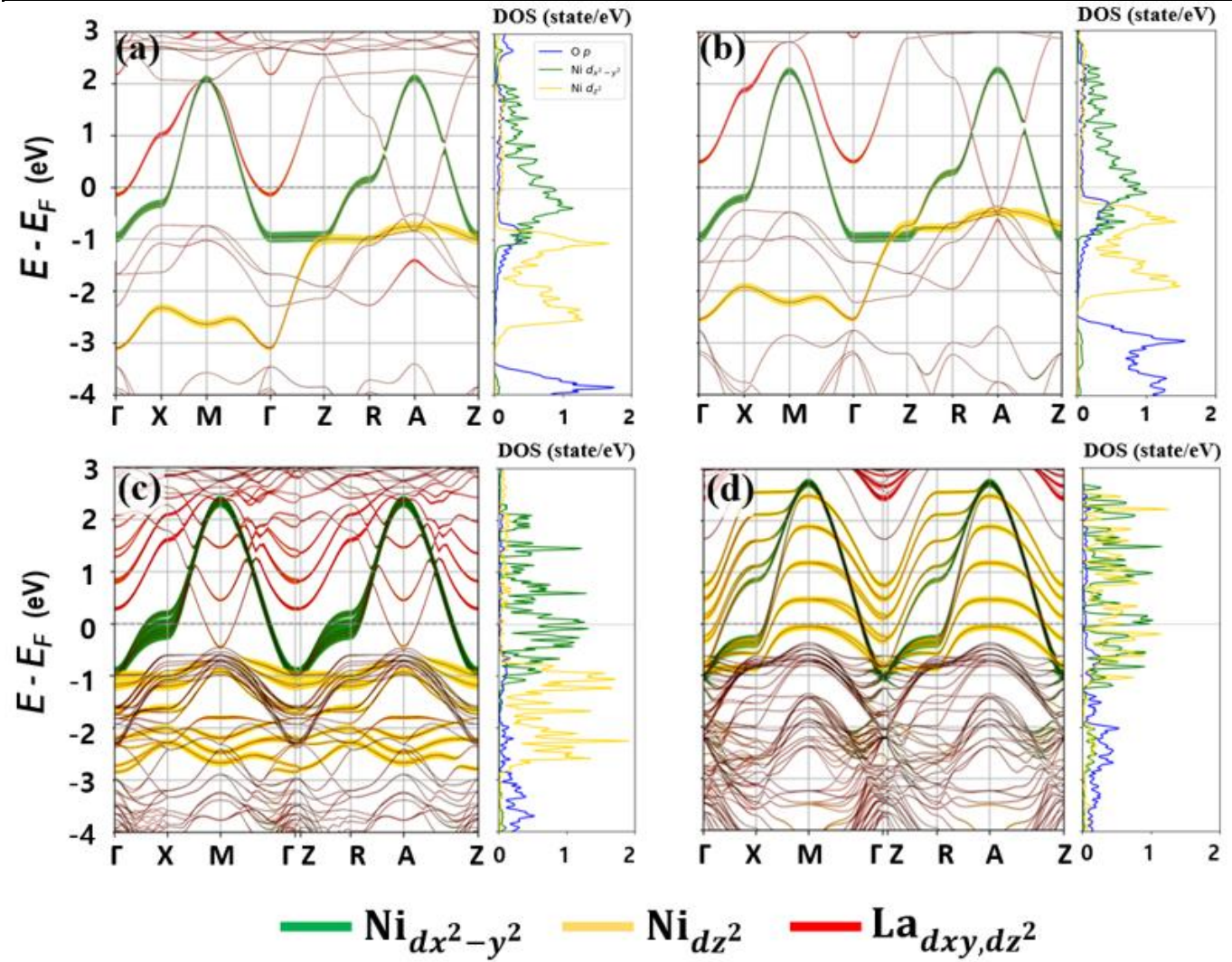

**Figure 2.** The calculated (nonmagnetic) band structure and PDOS for (a) the undoped ($\delta = 0$), (b) the hole-doped ($\delta = 0.2$) La112, (c) La6512 ($\delta = 0$), and (d) La6516 ($\delta = 0$). The green, yellow, and red colors denote the projected portion of Ni-$d_{x^2-y^2}$, Ni-$d_{z^2}$, and La-$d_{z^2}$, $d_{xy}$ states, respectively. In the DOS plots (right panels), oxygen states are also presented with blue lines while Ni-$d_{x^2-y^2}$ and Ni-$d_{z^2}$ is shown by green and yellow, respectively. Different types of oxygens in each system were averaged and then normalized for clarity of presentation. The more detailed features of O states are presented in Fig. 3.

In a notable recent work, Pan et. al. showed that thin film quintuple-layer nickelate Nd6512 becomes superconducting. [46]. In contrast to the previous case of NdNiO$_2$ (Nd112) for which rare-earth dopings (nominally corresponding to hole-dopings) induces superconducting transition, Nd6512 becomes superconducting without chemical doping. In fact, nominal charge counting gives rise to $d^{8.8}$ for R6512. Together with its structural similarity, Nd6512 has thus been regarded as the equivalent to the naturally hole-doped R112 [46,55].

Figure 2(a)-(d) presents the calculated band structure of undoped ($\delta = 0$), doped ($\delta = 0.2$) La112, the undoped La6512, and La6516, respectively. From the comparison of Fig. 2(a) with (c), the major band of Ni-$d_{x^2-y^2}$ character shows quite

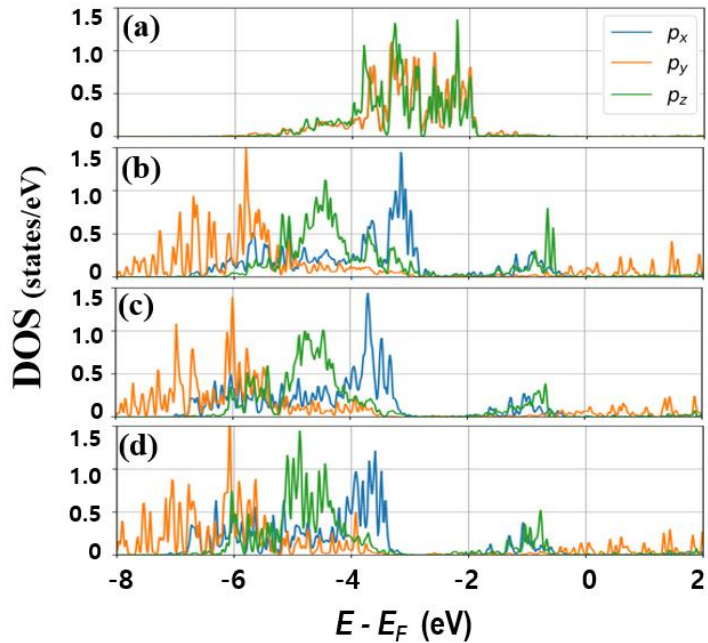

**Figure 3.** The layer-dependent projected DOS for oxygen in La6512. Oxygens in the (a) R2O2 layer, (b) 3rd-, (c) 2nd-, and (d) 1st-layer of NiO$_2$. O-$p_x$, $p_y$, and $p_z$ states are represented by sky-blue, orange, and green color, respectively.

similar dispersion in both systems. The noticeable band splitting only found in La6512 along Γ to X and Z to R is indicative of non-

  negligible interlayer hopping in quintuple layers as discussed earlier. A parabolic band of La-d$_{xy}$ character crosses Fermi level (E$_F$) at around A point in La112 while it is observed at both M ($\frac{1}{2},\frac{1}{2},0$) and A ($\frac{1}{2},\frac{1}{2},\frac{1}{2}$) in La6512. In n = 5 nickelate, a band complex of mostly Ni$_{t2g}$ character is noticed below E$_F$. Another intriguing similarity is observed in the La-d band. In the undoped La112, this band slightly touches E$_F$, forming a small electron pocket at Γ point; see the red-colored band in Fig.2(a). This pocket disappears by hole doping simply because of the lowered E$_F$ as shown in Fig. 2(b). In La6512 (Fig. 2(c)), the corresponding band locates well above the E$_F$ without chemical doping [63]. It is reminiscent of the formal valence difference between these two materials.

   It is instructive to compare the band structure of La$_6$Ni$_5$O$_{12}$ (Fig. 2(c)) with that of La$_6$Ni$_5$O$_{16}$ (Fig. 2(d)). As expected [57], the major changes are found in Ni-d$_{z2}$ bands (yellow-colored) whose energetic position is clearly lower and band width smaller in the case of La$_6$Ni$_5$O$_{12}$ due to the absence of apical anions. In La$_6$Ni$_5$O$_{16}$, the more bands go across the Fermi level, leading to a fairly different Fermi surface shape. Importantly, the d$_{x2-y2}$ states are distinctive as well. For example, the E$_F$-touching portion of bands at around X and R in Fig. 2(c) are found in the lower energy in Fig. 2(d). Another important difference is found in La states. The parabolic electron pocket at around M and A point in the case of La$_6$Ni$_5$O$_{12}$ (see Fig.2(c); corresponding to A point in Fig. 2(a)) is not noticeable in La$_6$Ni$_5$O$_{16}$. The absence of this 'self-doping' band can therefore be important to understand the induced superconductivity.

   Also intriguing is to note the O-p states, particularly their layer dependence. Fig. 3(b)-(d) shows the projected density of states (PDOS) for oxygen in the 3rd-, 2nd-, and 1st-NiO$_2$ layer, respectively. In the NiO$_2$ plane being closer to the R$_2$O$_2$

2024.03.13.

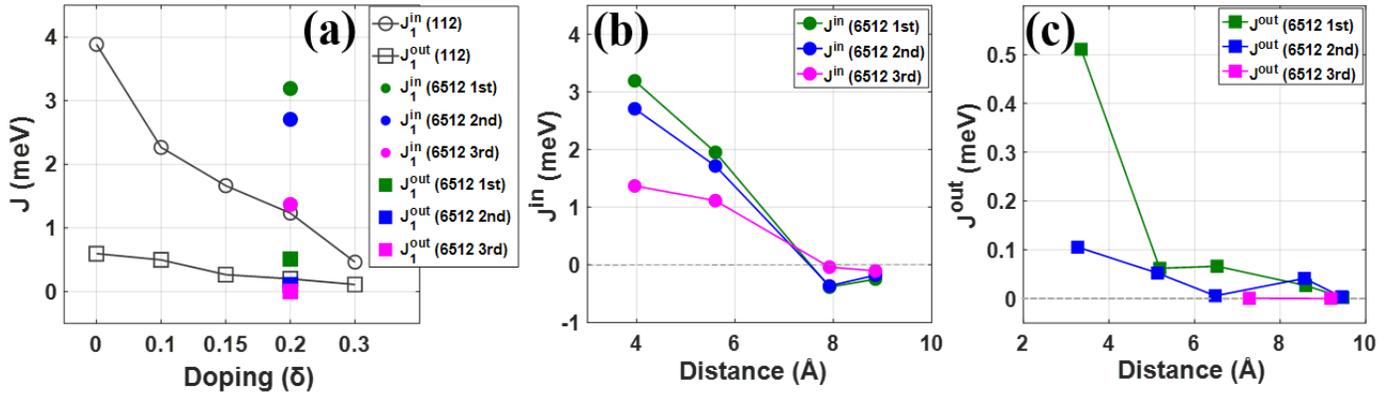

**Figure 4.** The calculated $J_1^{in}$ and $J_1^{out}$ of (undoped) La6512 (plotted at $\delta$=0.2 line of 112) in comparison with La112 (as a function of hole doping $\delta$). The calculation data of La112 are connected by gray line for clarity and expressed by open-circle ($J_1^{in}$) and open-square ($J_1^{out}$). For the quintuple-layer La6512, the results of three distinctive layers are presented with different colors. The green, blue and magenta represents the first, second, and third layer, respectively. The circles and squares show the in-plane and the out-of-plane magnetic couplings, respectively. (b, c) The calculated (b) in-plane ($J_n^{in}$) and (c) out-of-plane ($J_n^{out}$) couplings for La6512 as a function of Ni-Ni distance. The results of the first-, second-, and third-layer are depicted in green, blue, and magenta color, respectively.

blocking slab, the O states gradually move upward to $E_F$, demonstrating the role of fluorite slab as a hole doping source. It is also consistent with the calculation results of layer-dependence charges (see below). As expected, the oxygen states in $R_2O_2$ layer (Fig.3(a)) is very different from those in $NiO_2$.

Now let us turn our attention to magnetism. In particular, it is important to understand how the aforementioned similarities and differences in the crystal and the electronic structure between 112 and 6512 are manifested in their magnetic property. However, the underlying magnetism in R6512 has been less investigated although it can be responsible for superconducting pairing in this family of materials just as in cuprates While the magnetic interaction profile, for example, has been extensively studied for R112 [50,52,53], R6512 has never been investigated to the best of our knowledge. It is presumably because of the structural complexity of this quintuple-layer compound and the related difficulty in experiments and computations. The detailed quantitative and comparative understanding can provide a crucial insight to elucidate nickelate superconductivity.

First, we examine the relative stability of possible magnetic orders by considering several spin configurations within $\sqrt{2} \times \sqrt{2} \times 2$ unit cell (*Cmmm* space group). It should be noted that these orderings could not be observed in experiments even though they are well stabilized within our GGA computation scheme. It is indeed also the case for infinite-layer nickelates except that, for La112, there is a paper reporting c-type AFM ground state based on SCAN (strongly constrained and appropriately normed) DFT computation [54]. Just as in the previous studies for La112 [50,52,53], it is important to investigate the magnetic order and the couplings within this static snapshot picture of possibly fluctuating AFM spins as it gives rise to an appropriate approximation [50,52,53,64,65]. The calculated total energy of La6512 is found to be lowest for G-type AFM order. FM, A-type AFM, and C-type AFM phase has the greater energy than G-type by 50, 53, and 44 meV/Ni, respectively. Our result of FM versus G-type AFM order is consistent with the previous report [54]. Experimentally, there is no report of identifying the magnetic order while the FM order may be easily excluded [46]. Considering our results together with previous works, the possibility of the G-type antiferromagnetically ordered phase (without any defect) might not be excluded yet. Or, it is presumably more likely that the system is in the close vicinity to it with AFM fluctuation carrying no static order as observed in 112 nickelate and lightly-doped cuprates [5,8,10,35,36,50,52,53].

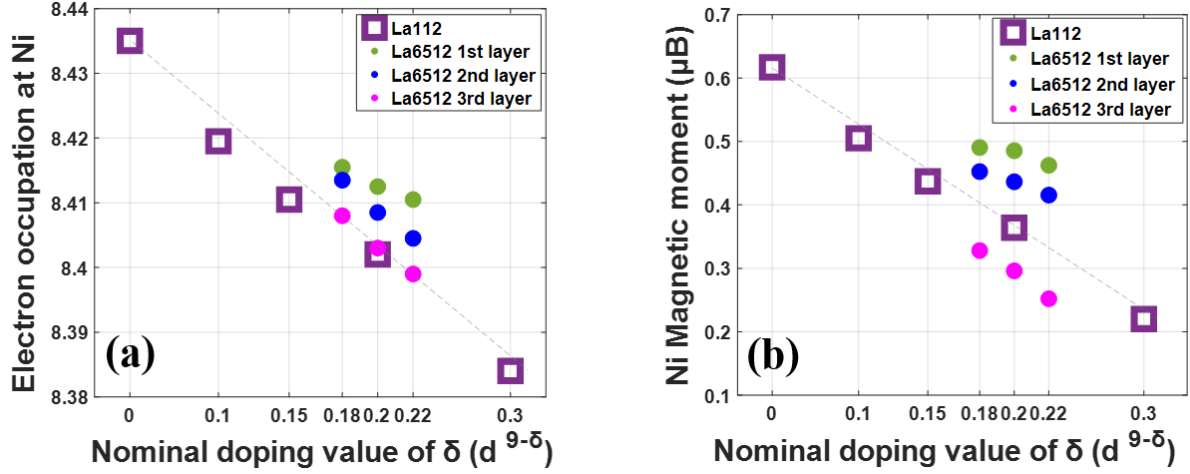

**Figure 5.** The calculated Ni electron occupations (a) and the magnetic moments (b) of La6512 (presented at $\delta$=0.18, 0.2 and 0.22 line) in comparison with La112 (plotted as a function of hole doping $\delta$). The results of La112 are depicted in purple open square. The green, blue, and magenta circles refer to the results of the first-, second-, and third- layer of La6512, respectively.

Based on MFT, we estimated magnetic interactions in La6512. There are three inequivalent layers as highlighted in Fig.1(d) and (e). Note that the inter-atomic distance corresponding to the out-of-plane coupling $J_1^{out}$ of the third layer is quite large (7.082 Å) and therefore it becomes very small; less than $10^{-4}$ meV (see, for example, Fig. 4(a) and (c)). Figure 4 summarizes our calculation results of the n-th neighboring in-plane ($J_n^{in}$) and the out-of-plane interactions ($J_n^{out}$). In Fig. 4(a), we also present the results of doped and undoped La112 for comparison, and the results of La6512 are plotted along the line of $\delta$=0.2 by considering its nominal charge. The decreasing feature of $J$ as a function of hole-doping level was noticed in La112 [50]. The gray-open circles and the open squares refer to $J_1^{in}$ and $J_1^{out}$, respectively, in La112 as a function of doping level. The filled-colored circles of green, blue, and magenta represent $J_1^{in}$ of the first, second, and third layer in La6512, respectively. The green-, blue-, and magenta-colored squares represent $J_1^{out}$ of the first, second, and third layer of La6512, respectively.

First of all, it is noted that the major magnetic coupling $J_1^{in}$ is clearly layer-dependent. As shown in Fig.4(a), the value corresponding to the first and second layer is found to be 3.19 and 2.70 meV, respectively. On the other hand, it is significantly reduced to be $J_1^{in}$ =1.37 meV for the third layer directly adjacent to the fluorite blocking layers. As discussed below, this more than factor of two reduction is closely related to the charge transfer. Within the simple ionic picture, introducing more holes can make the AFM superexchange weaker by pushing the system away from the nominal half-fillng of $d_{x2-y2}$ (corresponding to Ni-$d^9$).

Another intriguing finding is that the third layer $J_1^{in}$ value (filled magenta circle in Fig.4(a)) is in good agreement with $J_1^{in}$ of La112 (black open circle) at $\delta$=0.2. On the other hand, the inner-most layer $J_1^{in}$ (filled green circle) is found in between $\delta$=0 and $\delta$=0.1 of La112 values. This notable layer dependence of major in-plane coupling strength is attributed to the layer-different charge transfer (see below). The second layer $J_1^{in}$ (filled blue circle) is found in between the values of the first and the third layer. Our results on one hand indicate that the major magnetic interaction, namely $J_1^{in}$, of La6512 is consistent with the simple picture of regarding it as an appropriately doped case of La112 as discussed in literature [54,55]. Quantitatively, on the other, they clearly show the layer dependence. It can be useful information to understand the possibly existing intertwined orders in quintuple layer nickelate and their coherences distinctive from infinite layer [9].

As mentioned above, the layer-dependent magnetic interaction profile can be understood from the layer-dependent charge transfer. The calculated Ni electron occupation and the moment are summarized in Fig. 5(a) and (b), respectively. It is noted that our value of Ni-d occupation ~8.4 is smaller than the nominal one. As well known, the difference arises from the significant hybridization with neighboring O-p orbitals on top of the limitation in the charge counting methods [66,67]. Once again, the results of La6512 is shown at δ=0.2 line for comparison with La112. The Ni electron occupation in the outer-most (i.e., the third) $NiO_2$ layer (filled magenta circle) is markedly similar with that of La112 (empty squares) at δ=0.2 whereas the corresponding value increases as we move to the inner layers. See the filled green and the blue circle. Given that the $R_2O_2$ layer provides holes into $NiO_2$, the outer layer is expected to get more hole doped. This feature is also maintained when we introduce either holes or electrons into La6512 through VCA simulation; see the circular data points at δ=0.18 and 0.22 for hole and electron doping, respectively [68]. The calculated moment exhibits the same tendency; see Fig. 5(b). A detailed quantitative comparison shows an interesting difference; in terms of charge transfer amount, the first and the second layer Ni valences and moments correspond to the value of δ=0.1-0.15 of La112. In term of magnetic coupling $J_1^{in}$, on the other hand, it is in the range of $0 < δ < 0.1$ as can be seen in Fig.4(a). Naturally, the inter-site magnetic interaction depends on the electronic structure details.

Fig. 4(b) and (c) presents all magnetic couplings calculated as a function of inter-atomic Ni-Ni distance. It is found that aforementioned trend of $J^{in}$ (1st layer) > $J^{in}$ (2nd layer) > $J^{in}$ (3rd layer) is basically retained over a wide range of distance. As for $J^{out}$, the values are much smaller due to the longer distance and the third-layer $J^{out}$ is not quite meaningful (magenta colors in Fig. 4(c)) while it is clear that $J_1^{out}$(1st layer) > $J_1^{out}$(2nd layer).

## IV. SUMMARY AND CONCLUSION

In summary, we performed the first-principles calculations of magnetic interactions for the recently discovered superconducting quintuple-layer nickelate. As discussed in the previous studies, the system can be regarded as the doped case of another nickelate superconducting materials, namely, the infinite-layer nickelate from the point of view of electronic structure. Due to the different layer structure, however, the distribution of the transferred charges from the fluorite layers are naturally layer-dependent, and it results in the noticeable layer dependence of the magnetic interactions in La6512. While the outer most $NiO_2$ layer has the major in-plane exchange coupling well comparable with the doped La112 at nominal valence of δ=0.2, it gradually becomes stronger and approaches toward the undoped value of δ=0 as we goes into the inner layers. The analysis on the layer-dependent charge occupation and magnetic moment are consistent with the MFT results and thus support our interpretation. The tendency of the greater coupling in the inner layer is maintained up to the quite long range. We also present the out-of-plane couplings over a wide range. Our results provide the useful information to understand the quintuple layer nickelate and its superconductivity in comparison to other related materials. Recently, there have been several reports of new Ni-based superconducting materials such as $La_3Ni_2O_7$ and $La_4Ni_3O_{10}$ [69-72]. For the former, in particular, notable progress has been made in understanding its distinctive characteristics from infinite- and quintuple-layer nickelates. According to recent theoretical studies, this new family carrying a formal valence of +2.5 can be a bilayer Hubbard system in which the interlayer interaction plays an important role [73-79].

**ACKNOWLEDEMENTS**

Authors are grateful to Siheon Ryee for helpful discussion. This work was supported by the National Research Foundation of Korea (NRF) grant funded by the Korea government (MSIT) (Grant Nos. 2021R1A2C1009303 and RS-2023-00253716)